# Transivity of Commutativity for Second-Order Linear Time-Varying Analog Systems

Mehmet Emir KOKSAL

*Department of Mathematics, Ondokuz Mayis University, 55139 Atakum, Samsun, Turkey*

*emir_koksal@hotmail.com*

**Abstract:** After reviewing commutativity of second-order linear time-varying analog systems, the inverse commutativity conditions are derived for these systems by considering non-zero initial conditions. On the base of these conditions, the transitivity property is studied for second order linear time-varying unrelaxed analog systems. It is proven that this property is always valid for such systems when their initial states are zero; when non-zero initial states are present, it is shown that the validity of transitivity does not require any more conditions and it is still valid. Throughout the study it is assumed that the subsystems considered can not be obtained from each other by any feed-forword and feed-back structure. The results are well validated by MATLAB simulations.

## I. Introduction

Second-order differential equations originate in electromagnetic, electrodynamics, transmission lines and communication, circuit and system theory, wave motion and distribution, and in many fields of electrics-electronics engineering. They play a prominent role for modelling problems occurring in electrical systems, fluid systems, thermal systems and control systems. Especially, they are used as a powerful tool for modelling, analyzing and solving problems in classical control theory, modern control theory, robust control theory and automatic control, which is essential in any field of engineering and sciences, and for discussing the results turned up at the end of analyzing for resolution of naturel problems. For example, they are used in cascade connected and feedback systems to design higher order composite systems for achieving several beneficial properties such as controllability, sensitivity, robustness, and design flexibility. When the cascade connection which is an old but still an up to date trend in system design [1-4] is considered, the commutativity concept places an important role to improve different system performances. On the other hand, since the commutativity of linear time-invariant relaxed systems is straightforward and time-varying systems have found a great deal of applications recently [5-10], the scope of this paper is focused on commutativity of linear time-varying systems only.

When two systems *A* and *B* are are interconnected one after the other so that the output of the former acts as the input of the later, it is said that these systems are connected in cascade [11]. If the order of connection in the sequence is not effective on the input-output relation of the cascade connection, then these systems are commutative [12].

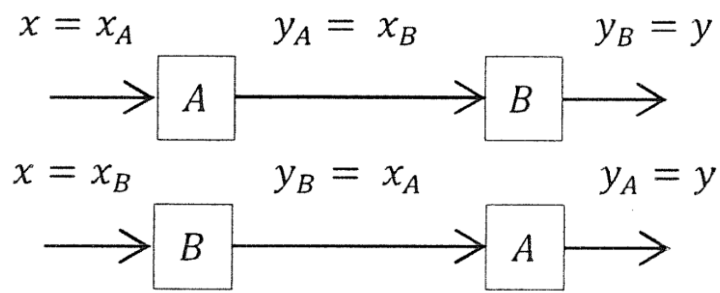

**Figure 1:** Cascade connection of the differential system *A* and *B*

The tutorial papers [13, 14] cover almost all the subjects scattered in a great deal of literature about commutativity. Some of the important results about the commutativity are summarized in the sequel superficially.

J. E. Marshall has proven that "for commutativity, either both systems are time-invariant or both systems are time-varying" [12]. After many contributions appeared as conference presentations and a few short papers focusing to special cases such as first, second, third, and forth order systems, the exhaustive journal paper of M. Koksal introduced the basic fundamentals of the subject [13]. Another work joint by the same author has presented explicit commutativity conditions of fifth order systems in addition to reviews of commutativity of systems with non-zero initial conditions, commutativity and system disturbance, commutativity of Euler systems [14].

There is some literature on the commutativity of discrete-time systems as well [15, 16]. And the research of commutativity is continuing on both analog and digital systems in [17, 18]. In [17], all the second-order commutative pairs of a first-order linear time-varying analogue systems are derived. The decomposition of a second-order linear time-varying systems into its first-order commutative pairs are studied in [18]. This is important for the cascade realization of the second-order linear time-varying systems.

In [19], the inverse conditions expressed in terms of the coefficients of the differential equation describing system **B** have been derived for the case of zero initial conditions and shown to be of the same form of the original equations appearing in the literature.

Transitivity property of commutativity is first introduced and some general conditions for it are presented in [20], where transitivity of commutativity is fully investigated for first-order systems. It is shown that commutativity of first-order linear time-varying systems with and without initial conditions has always transitivity property.

Explicit commutativity conditions for second-order linear time-varying systems have been studied in [21]. On the other hand, no special research as in [20] for the first-order systems has been done on the transitivity property of commutativity for second-order systems. This paper completes this vacancy in the literature on the base of work in [21].

In this paper, transtivity property of second-order linear time-varying analog systems with and without initial conditions is studied. Section II is devoted to the explicite commutativity conditions for such systems. Section III presents preliminaris, namely inverse commutativity conditions in the case of non-zero inititial states, that are used in the proof of the subseqient section. Section IV deals with transtivity property with and without initial conditions. After giving an example is Section V, the paper ends with conclusions which appears in Section VI.

**II.     Commutativity Conditions for Second-order Systems**

In this section, the commutativity conditions for two second-order linear time-varying analog systems are reviewed [21].

Let $A$ be the systam described by the second-order linear time-varying differential equation

$$a_2(t)\ddot{y}_A(t) + a_1(t)\dot{y}_A(t) + a_0(t)y_A(t) = x_A(t); t \geq t_0 \tag{1a}$$

with the initial conditions at the initial time $t_0 \in R$

$$y_A(t_0), \dot{y}_A(t_0). \tag{1b}$$

Where the single (double) dot on the top indicates the first (second) order derivative with respect to time $t \in R$; $x_A(t)$ and $y_A(t)$ are the input and output of the system, respectively. Since the system is second-order

$$a_2(t) \not\equiv 0. \tag{1c}$$

Further, $\ddot{a}_2(t), \dot{a}_1(t)$ and , $a_0(t)$ are well defined continuous functions, that is $a_2(t)$ , $\dot{a}_1(t)$, $a_0(t) \in C[t_0, \infty)$ , hence so $\dot{a}_2(t), a_2(t), a_1(t)$ are.

It is true that System $A$ has a unique continuous solution $y_A(t)$ with its first and second-order derivatives for any continuous input function $x_A(t)$ [22].

Let $B$ be another second-order linear time varying system described in a similar way to $A$. Hence it is described by

$$b_2(t)\ddot{y}_B(t) + b_1(t)\dot{y}_B(t) + b_0(t)y_B(t) = x_B(t), t \geq t_0, \quad (2a)$$

$$y_B(t_0), \dot{y}_B(t_0), \quad (2b)$$

$$b_2(t) \not\equiv 0. \quad (2c)$$

Where the coefficients $b_2, b_1, b_0$ satisfy the same properties satisfied by $a_2, a_1, a_0$.

For the commutativity of $B$ with $A$, it is well known that

I)     i) The coefficients of $B$ be expressed in terms of those of $A$ through

$$\begin{bmatrix} b_2(t) \\ b_1(t) \\ b_0(t) \end{bmatrix} = \begin{bmatrix} a_2(t) & 0 & 0 \\ a_1(t) & a_2^{0.5}(t) & 0 \\ a_0(t) & f_A(t) & 1 \end{bmatrix} \begin{bmatrix} k_2 \\ k_1 \\ k_0 \end{bmatrix}, \quad (3a)$$

where

$$f_A(t) = \frac{a_2^{-0.5}[2a_1(t) - \dot{a}_2(t)]}{4}. \quad (3b)$$

And $k_2, k_1, k_0$ are some constants with $k_2 \neq 0$ for $B$ is of second-order. Further, it is assumed that $B$ can not be obtained from $A$ by a constant feed forward and feedback path gains; hence $k_1 \neq 0$ [14, 21].

    ii)                 $a_0 - f_A^2 a_2^{-0.5}\dot{f}_A = A_0, \forall t \geq t_0,$            (3c)

where $A_0$ is a constant,

When the initial conditions in (1b) and (2b) are zero, the conditions i) and ii) are necessary and sufficient conditions fort the commutativity of $A$ and $B$ under the mentioned conditions ($k_2 \neq 0, k_1 \neq 0$).

For commutativity with nonzero initial conditions as well, the following additional conditions are required:

II)     i)

$$y_B(t_0) = y_A(t_0) \neq 0, \tag{4a}$$

$$\dot{y}_B(t_0) = \dot{y}_A(t_0); \tag{4b}$$

ii)

$$(k_2 + k_0 - 1)^2 = k_1^2(1 - A_0), \tag{4c}$$

iii)

$$\dot{y}_B(t_0) = -a_2^{-0.5}(t_0)\left[\frac{k_2+k_0-1}{k_1} + f_A(t_0)\right]y_B(t_0); \tag{4d}$$

which are necessary and sufficient together with Eqs. 3a,b,c for commutativity of $A$ and $B$ under non-zero initial conditions. Note that by the nonzero initial condition it is meant "general values of initial conditions", so one or two of them may be zero in special cases. In fact, if the output $y(t_0)$ is zero, its derivative need to be zero due to Eq. 4d; if not, its derivative may or may not be zero depending on the term in bracket in (4d) is zero or not [21].

### III. Inverse Commutativity Conditions for Second Order Unrelaxed Systems

For the proof of the transitivity theorems of the following section, we need some formulas which express the inverse commutativity conditions. Although these conditions have been partially treated in [21], initial conditions are all assumed to be zero their; and for the sake of completeness, we express the general results by Lemma 1 and exhibit the complete inverse commutativity conditions for unrelaxed second order linear time-varying systems. In the previous section, the necessary and sufficient conditions for the commutativity of $A$ and $B$ are expressed dominantly by awnsening "what are the conditions that must be satisfied by $B$ to be commutative with $A$?" The answer to this question constitutes the inverse commutativity conditions. We express the results by the following Lemma.

**Lemma 1.** *The necessary and sufficients conditions given in Eqs. (3) and (4) for the commutativity A and B can be expressed by the following formulas:*

I.    i)

$$\begin{bmatrix} a_2(t) \\ a_1(t) \\ a_0(t) \end{bmatrix} = \begin{bmatrix} b_2(t) & 0 & 0 \\ b_1(t) & b_2^{0,5}(t) & 0 \\ b_0(t) & f_B(t) & 1 \end{bmatrix} \begin{bmatrix} l_2 \\ l_1 \\ l_0 \end{bmatrix}, \text{ where} \quad (5a)$$

$$f_B(t) = \frac{b_2^{-0,5}[2b_1(t) - \dot{b}_2(t)]}{4}. \quad (5b)$$

ii)

$$b_0 - f_B^2 - b_2^{0,5} \dot{f}_B = B_0, \forall t \geq t_0 \quad (5c)$$

II.  i)

$$y_A(t_0) = y_B(t_0), \quad (6a)$$

$$\dot{y}_A(t_0) = \dot{y}_B(t_0); \quad (6b)$$

ii)

$$(l_2 + l_0 - 1)^2 = l_1^2(1 - B_0), \quad (6c)$$

$$\dot{y}_A(t_0) = -b_2^{-0.5}(t_0)\left[\frac{l_2 + l_0 - 1}{l_1} + f_B(t_0)\right] y_A(t_0); \quad (6d)$$

For the proof of the lemma, we solve Eq. 3a for $a_2, a_1, a_0$ in terms of $b_2, b_1, b_0$ as follows:

$$a_2(t) = \frac{1}{k_2} b_2(t), \quad (7a)$$

$$a_1(t) = \frac{b_1(t) - k_1 a_2^{0.5}(t)}{k_2} = \frac{1}{k_2} b_1(t) - \frac{k_1}{k_2}\left(\frac{b_2}{k_2}\right)^{0,5} = \frac{1}{k_2} b_1(t) - \frac{k_1}{k_2^{1,5}} b_2^{0.5}(t). \quad (7b)$$

Before progressing further, let us compute $f_A$ from (3b) and by using the above formulas for $a_2$ and $a_1$, we obtain

$$f_A = \frac{1}{4} \frac{b_2^{-0.5}}{k_2}\left[2\left(\frac{b_1}{k_2} - \frac{k_1 b_2^{0.5}}{k_2^{1.5}}\right) - \frac{\dot{b}_2}{k_2}\right] = \frac{1}{4} \frac{b_2^{-0.5}}{k_2^{-0.5}}\left[\frac{2b_1 - \dot{b}_2}{k_2} - \frac{2k_1 b_2^{0.5}}{k_2^{1.5}}\right]$$

$$= k_2^{-0.5} \frac{b_2^{-0.5}(2b_1 - \dot{b}_2)}{4} - \frac{k_1}{2k_2}$$

Finally, defining $b_2^{-0.5}(2b_1 - \dot{b}_2)/4$ as $f_B$ as in (5b), we have

$$f_A = k_2^{-0.5} f_B - \frac{k_1}{2k_2}, \quad (8a)$$

or equivalently

$$f_B = k_2^{0.5} f_A + \frac{k_1}{2k_2^{0.5}} \tag{8b}$$

We now compute $a_0$ from the last row of (3a) by using (8a)

$$a_0 = \frac{b_0 - k_1 f_A - k_0}{k_2} = \frac{b_0}{k_2} - \frac{k_1}{k_2}\left[k_2^{-0.5} f_B + \frac{k_1}{2k_2}\right] - \frac{k_0}{k_2}$$

$$= \frac{1}{k_2} b_0 - \frac{k_1}{k_2^{1.5}} f_B + \frac{k_1^2}{2k_2^2} - \frac{k_0}{k_2}. \tag{9}$$

Writing (7a), (7b), and (9) in matrix form we obtain

$$\begin{bmatrix} a_2 \\ a_1 \\ a_0 \end{bmatrix} = \begin{bmatrix} b_2 & 0 & 0 \\ b_1 & b_2^{0.5} & 0 \\ b_0 & f_B & 1 \end{bmatrix} \begin{bmatrix} \frac{1}{k_2} \\ \frac{-k_1}{k_2^{1.5}} \\ \frac{k_1^2}{2k_2^2} - \frac{k_0}{k_2} \end{bmatrix}, \tag{10}$$

Comparing with Eq. 5a, we observe that (5a) is valid with

$$\begin{bmatrix} l_2 \\ l_1 \\ l_0 \end{bmatrix} = \begin{bmatrix} \frac{1}{k_2} \\ \frac{-k_1}{k_2^{1.5}} \\ \frac{k_1^2}{2k_2^2} - \frac{k_0}{k_2} \end{bmatrix}. \tag{11a}$$

Hence (5a) has been proved.

For use in the sequel, we solve (11a) for $k_i$'s and obtain

$$\begin{bmatrix} k_2 \\ k_1 \\ k_0 \end{bmatrix} = \begin{bmatrix} \frac{1}{l_2} \\ \frac{-l_1}{l_2^{1.5}} \\ \frac{l_1^2}{2l_2^2} - \frac{l_0}{l_2} \end{bmatrix}, \tag{11b}$$

which is naturally the dual of Eq. 11a with $k$ and $l$ iterchanged. By using (11b) in (8a) and (8b), or directly interchanging $A \leftrightarrow B$ and $k_i \leftrightarrow l_i$ in (8a) and (8b), we obtain the following equations:

$$f_B = l_2^{-0.5} f_A - \frac{l_1}{2l_2}, \tag{11c}$$

$$f_A = l_2^{0.5}\left(f_B + \frac{l_1}{2l_2}\right). \tag{11d}$$

To show (5c), we substitute values of $b_2$ and $b_0$ from (3a), value of $f_B$ from (8b) in the left side of (5c), we obtain

$$b_0 - f_B^2 - b_2^{0.5}\dot{f}_B = k_2 a_0 + k_1 f_A + k_0 - (k_2^{0.5} f_A + \frac{k_1}{2k_2^{0.5}})^2 - (k_2 a_2)^{0.5}(k_2^{0.5}\dot{f}_A)$$

$$= k_2 a_0 + k_1 f_A + k_0 - k_2 f_A^2 - k_1 f_A - \frac{k_1^2}{4k_2} - k_2 a_2^{0.5}\dot{f}_A$$

$$= k_2(a_0 - f_A^2 - a_2^{0.5}\dot{f}_A) + k_0 - \frac{k_1^2}{4k_2}.$$

Finally using (3c)

$$b_0 - f_B^2 - b_2^{0.5}\dot{f}_B = k_2 A_0 + k_0 - \frac{k_1^2}{4k_2}$$

which is constant for $A_0$ being constant. Hence (5c) is valid with

$$B_0 = k_2 A_0 + k_0 - \frac{k_1^2}{4k_2}, \tag{12a}$$

or equivalently

$$A_0 = \frac{1}{k_2} B_0 + \frac{k_0}{k_2} + \frac{k_1^2}{4k_2}. \tag{12b}$$

The dual equations for (12a) and (12b) can be written by using constants $l_i$'s; this is done by using (11b) in (12a) and (12b), or directly interchanging $A \leftrightarrow B$ and $k_i \leftrightarrow l_i$ in (12a) and (12b). The results are

$$A_0 = l_2 B_0 + l_0 - \frac{l_1^2}{4l_2}, \tag{12c}$$

$$B_0 = \frac{1}{l_2} A_0 - \frac{l_0}{l_2} + \frac{l_1^2}{4l_2^2}. \tag{12d}$$

Equations (6a) and (6b) are the same as Eqs. 4a and 4b, respectively, so they do not need to be proved.

To prove (6c), we start from (4c); inserting valves of $k_i$'s from (11b) and valve of $A_0$ from (12b) in, we obtain:

$$(\frac{1}{l_2} + \frac{l_1^2}{2l_2^2} - \frac{l_0}{l_2} - 1)^2 = \frac{l_1^2}{l_2^3}(1 - l_2 B_0 - l_0 + \frac{l_1^2}{4l_2})$$

$$(\frac{1-l_0-l_2}{l_2}+\frac{l_1^2}{2l_2^2})^2 = \frac{l_1^2}{l_2^3}(1-l_2B_0-l_0+\frac{l_1^2}{4l_2})$$

$$(1-l_0-l_2+\frac{l_1^2}{2l_2})^2 = \frac{l_1^2}{l_2}(1-l_2B_0-l_0+\frac{l_1^2}{4l_2})$$

$$(1-l_0-l_2)^2 + \frac{l_1^4}{4l_2^2} + \frac{l_1^2}{l_2}(1-l_0-l_2) = \frac{l_1^2}{l_2}((1-l_2B_0-l_0)+\frac{l_1^4}{4l_2^2})$$

$$(1-l_0-l_2)^2 = \frac{l_1^2}{l_2}(1-l_2B_0-l_0-1+l_0+l_2)$$

$$(l_2+l_0-1)^2 = l_1^2(1-B_0)$$

To prove (6d), we start from (4d) and using (4a) and (4b), we write

$$\dot{y}_A(t_0) = -a_2^{-0.5}(t_0)\left[\frac{k_2+k_0-1}{k_1}+f_A(t_0)\right]y_A(t_0).$$

Using (5a) for $a_2$, (11b) for $k_i$'s and (11d) for $f_A$, we proceed

$$\dot{y}_A(t_0) = -[l_2b_2(t_0)]^{-0.5}\left[\frac{\frac{1}{l_2}+\frac{l_1^2}{2l_2^2}-\frac{l_0}{l_2}-1}{-\frac{l_1}{l_2^{1.5}}}+l_2^{0.5}f_B(t_0)+\frac{l_1}{2l_2^{0.5}}\right]y_A(t_0)$$

$$= -l_2^{-0.5}b_2^{-0.5}(t_0)[-\frac{l_2^{1.5}}{l_1}(\frac{1}{l_2}+\frac{l_1^2}{2l_2^2}-\frac{l_0}{l_2}-1)+l_2^{0.5}f_B(t_0)+\frac{l_1}{2l_2^{0.5}}]y_A(t_0)$$

$$= -b_2^{-0.5}(t_0)[-\frac{1}{l_1}-\frac{l_1}{2l_2}+\frac{l_0}{l_1}+\frac{l_2}{l_1}+f_B(t_0)+\frac{l_1}{2l_2}]$$

$$= -b_2^{-0.5}(t_0)[\frac{l_2+l_0-1}{l_1}+f_B(t_0)]y_A(t_0)$$

which is the same equation as (6d). Hence the proof of Lemma 1 is completed.

**Fact:** Comparing Eqs. 4d and 6d, together with the equalities $y_A(t_0) = y_B(t_0)$, $\dot{y}_A(t_0) = \dot{y}_B(t_0)$, we see that the derivatives $\dot{y}_A(t_0)$ and $\dot{y}_B(t_0)$ are constant multiples of $y_A(t_0)$ and $y_B(t_0)$. The multipliers are initial time dependent and

$$-ab_2^{-0.5}(t_0)\left[\frac{k_2+k_0-1}{k_1}+f_A(t_0)\right] = -b_2^{-0.5}(t_0)[\frac{l_2+l_0-1}{l_1}+f_B(t_0)].$$

Inserting in value of $b_2$ from (3a) and value of $f_B(t_0)$ from (8b) yields that

$$\frac{l_2+l_0-1}{l_1} = k_2^{0.5}[\frac{k_2+k_0-1}{k_1}-\frac{k_1}{2k_2}].$$

On the other hand, inserting in value of $a_2$ from (5a) and value of $f_A(t_0)$ from (11d) yields

$$\frac{k_2 + k_0 - 1}{k_1} = l_2^{0.5}\left[\frac{l_2 + l_0 - 1}{l_1} - \frac{l_1}{2l_2}\right].$$

This is the dual of the previous equation. Using the transformations (11a) and (11b) between $k_i$'s and $l_i$'s, it is straightforword to show that the above relations between $k_i$'s and $l_i$'s are valid.

## IV. Transitivity Property of Commutativity

To be able to study the transivity property of commutativity for second-order linear time-varying systems, we should consider a third system $C$ of the same type as $A$ and $B$ considered in Section III. So, let $C$ be defined by the following second-order differential equation:

$$c_2(t)\ddot{y}_C(t) + c_1(t)\dot{y}_C(t) + c_0(t)y_C(t) = x_C(t); t \geq t_0, \tag{13a}$$

$$y_C(t_0), \dot{y}_C(t_0), \tag{13b}$$

$$c_2(t) \not\equiv 0, \tag{13c}$$

where $\ddot{c}_2(t), \dot{c}_2(t), c_0(t) \in C[t_0, \infty)$. We assume $C$ is commutative with $B$, to similar relations to Eqs. (3) and (4) can be written as

I.  i)

$$\begin{bmatrix} c_2 \\ c_1 \\ c_0 \end{bmatrix} = \begin{bmatrix} b_2 & 0 & 0 \\ b_1 & b_2^{0.5} & 0 \\ b_0 & f_B & 1 \end{bmatrix} \begin{bmatrix} m_2 \\ m_1 \\ m_0 \end{bmatrix}, \text{where} \tag{14a}$$

$$f_B = \frac{b_2^{0.5}(2b_1 - \dot{b}_2)}{4}. \tag{14b}$$

And $m_2, m_1, m_0$ are some constants with $l_2 \neq 0$ for $C$ is of second-order. Further, we assume that $C$ can not be obtained from $B$ by constant feed forward and feedback gains; hence $m_1 \neq 0$. Moreover,

ii)

$$b_0 - f_B^2 - b_2^{0.5}\dot{f}_B = B_0, \quad \forall t \geq t_0 \tag{14c}$$

where $B_0$ is a constant. When the initial conditions are non-zero, the following should be satisfied:

II.  i)
$$y_C(t_0) - y_B(t_0) \neq 0, \tag{15a}$$
$$\dot{y}_C(t_0) = \dot{y}_B(t_0), \tag{15b}$$

ii)
$$(m_2 + m_0 - 1)^2 = m_1^2(1 - B_0), \tag{15c}$$
$$\dot{y}_C(t_0) = -b_2^{-0.5}\left[\frac{m_2+m_0-1}{m_1} + f_B(t_0)\right]y_C(t_0). \tag{15d}$$

Considering the inverse commutativity conditions derived for $A$ and $B$, the inverse commutativity conditions for $B$ and $C$ can be written from Eqs. (5) and (6) by changing $A \to B$ and $B \to C$, $h_i \to m_i$ and $l_i \to n_i$ in Eqs. (5) and (6). The results are

I.  i)
$$\begin{bmatrix} b_2 \\ b_1 \\ b_0 \end{bmatrix} = \begin{bmatrix} c_2 & 0 & 0 \\ c_1 & c_2^{0.5} & 0 \\ c_0 & f_c & 1 \end{bmatrix}\begin{bmatrix} n_2 \\ n_1 \\ n_0 \end{bmatrix}, \text{where} \tag{16a}$$

$$f_c = \frac{c_2^{-0.5}(2c_1-c_2)}{4}. \tag{16b}$$

ii)
$$c_0 - f_c^2 - c_2^{0.5}f_c = c_0, \forall t \geq t_0 \tag{16c}$$

II.  i)
$$y_B(t_0) = y_C(t_0), \tag{17a}$$
$$y'_B(t_0) = y'_C(t_0), \tag{17b}$$

ii)
$$(n_2 + n_0 - 1)^2 = n_1^2(1 - c_0), \tag{17c}$$
$$\dot{y}_B(t_0) = -c_2^{-0.5}(t_0)\left[\frac{n_2+n_0-1}{n_1} + f_c(t_0)\right]y_B(t_0). \tag{17d}$$

Further, Eqs, (8a) and (8b) become

$$f_B = m_2^{-0.5} f_C - \frac{2m_1}{m_2}, \tag{18a}$$

$$f_C = m_2^{0.5} f_B + \frac{m_1}{2m_2^{0.5}}. \tag{18b}$$

The relations between the constants $m_i$ and $n_i$ can be written from Eqs. (11a) and (11b) by the replacements $k_i \to m_i, l_i \to n_i$. The results are;

$$\begin{bmatrix} n_2 \\ n_1 \\ n_0 \end{bmatrix} = \begin{bmatrix} \frac{1}{m_2} \\ -\frac{m_1}{m_2^{1.5}} \\ \frac{m_1^2}{2m_2^2} - \frac{m_0}{m_2} \end{bmatrix}, \tag{19a}$$

$$\begin{bmatrix} m_2 \\ m_1 \\ m_0 \end{bmatrix} = \begin{bmatrix} \frac{1}{n_2} \\ -\frac{n_1}{n_2^{1.5}} \\ \frac{n_1^2}{2n_2^2} - \frac{n_0}{n_2} \end{bmatrix}. \tag{19b}$$

By using values of $m_i$'s in Eqs. (18a) and (18b), or directly interchanging $B \leftrightarrow C, m_i \leftrightarrow n_i$, we obtain

$$f_C = n_2^{-0.5} f_B - \frac{2n_1}{n_2}, \tag{19c}$$

$$f_B = n_2^{0.5} f_C + \frac{n_1}{2n_2^{0.5}}). \tag{19d}$$

Finally, Eqs. (12a, b, c, d) turns out to be

$$C_0 = m_2 B_0 + m_0 - \frac{m_1^2}{4m_2}, \tag{20a}$$

$$B_0 = \frac{1}{m_2} C_0 - \frac{m_0}{m_2} + \frac{m_1^2}{4m_2^2}, \tag{20b}$$

$$B_0 = n_2 C_0 + n_0 - \frac{n_1^2}{4n_2}, \tag{20c}$$

$$C_0 = \frac{1}{n_2} B_0 - \frac{n_0}{n_2} + \frac{n_1^2}{4n_2^2}. \tag{20d}$$

by the replacements $A \to B, B \to C, k_i \to m_i, l_i \to n_i$.

The preliminaries have been ready now for studying the transitivity property of commutativity. Assuming $B$ is commutative with $A$ and $C$ is commutative with $B$, we need to answer that weather $C$ is a commutative pair of $A$. The answer is expressed by the following theorems and their proves.

**Theorem 1:** *Transitivity property of commutativity for second-order linear time-varying analog systems which cannot be obtained from each other by constant feed forward and feedback gains is always valid under zero initial conditions.*

**Proof:** Since it is true by the hypothesis that $B$ is commutative with $A$, Eqs. (3a) and (3c) are valid; since it is true by hypothesis that $C$ is commutative with $B$, Eqs. (14a) and (14c) are also valid.

To prove Theorem 1, it should be proven that $C$ is commutative with $A$ under zero initial conditions. Referring to the commutativity conditions for $A$ and $B$ in Eq. (3a) and replacing $B$ by $C$, this proof is done by showing the validity of

$$\begin{bmatrix} c_2 \\ c_1 \\ c_0 \end{bmatrix} = \begin{bmatrix} a_2 & 0 & 0 \\ a_1 & a_2^{0,5} & 0 \\ a_0 & f_A & 1 \end{bmatrix} \begin{bmatrix} p_2 \\ p_1 \\ p_0 \end{bmatrix}, \qquad (20)$$

where $f_A(t)$ is given as in Eq. (3b), and the coefficients of $A$ already satisfy Eq. (3c) due to the commutativity of $B$ with $A$; further $p_2, p_1, p_0$ are some constants to be revealed:

Using Eq. (14a) first and then Eq. (3a), as well as Eq. (8b) for computing $c_0$, we can express $c_2, c_1, c_0$ as follows:

$$c_2 = m_2 b_2 = m_2 k_2 a_2,$$

$$c_1 = m_2 b_1 + m_1 b_2^{0,5} = m_2(k_2 a_1 + k_1 a_2^{0,5}) + m_1 (h_2 a_2)^{0,5}$$

$$= m_2 h_2 a_1 + (m_2 k_1 + m_1 k_2^{0,5}) a_2^{0,5}$$

$$c_0 = m_2 b_0 + m_1 f_B + m_0$$

$$= m_2(k_2 a_0 + k_1 f_A + k_0) + m_1 \left( k_2^{0,5} f_A + \frac{k_1}{2 k_2^{0,5}} \right) + m_0$$

$$= m_2 k_2 a_0 + (m_2 k_1 + m_1 k_2^{0,5}) f_A + m_2 k_0 + \frac{m_1 k_1}{2 k_2^{0,5}} + m_0.$$

These results can be written as

$$\begin{bmatrix} c_2 \\ c_1 \\ c_0 \end{bmatrix} = \begin{bmatrix} a_2 & 0 & 0 \\ a_1 & a_2^{0.5} & 0 \\ a_0 & f_A & 1 \end{bmatrix} \begin{bmatrix} m_2 k_2 \\ m_2 k_1 + m_1 k_2^{0.5} \\ m_2 k_0 + \frac{m_1 k_1}{2 k_2^{0.5}} + m_0 \end{bmatrix}, \qquad (21a)$$

which is exactly in the same form as Eq. (20) with the constants $p_2, p_1, p_0$;

$$\begin{bmatrix} p_2 \\ p_1 \\ p_0 \end{bmatrix} = \begin{bmatrix} m_2 k_2 \\ m_2 k_1 + m_1 k_2^{0.5} \\ m_2 k_0 + \frac{m_1 k_1}{2 k_2^{0.5}} + m_0 \end{bmatrix}. \qquad (21b)$$

So, the proof is completed.

For the validity of transitivity property for second-order linear time-varying analog systems under non-zero initial conditions, we state the following theorem.

**Theorem 2:** *Transitivity property of commutativity of systems considered in Theorem 1 is valid for the non-zero initial conditions of the systems as well.*

**Proof:** The proof is done by showing the commutativity of $C$ with $A$ under non-zero conditions as well. Since $C$ and $A$ are commutative with non-zero initial conditions Eq. (20) and Eq. (3a) are valid as mentioned in proof of Theorem 1. To complete the proof, we should show that $C$ is a commutative pair of $A$ under non-zero conditions as well, Eqs. (4a-d) are satisfied for systems $C$ (instead of $B$) and $A$. Namely,

$$y_C(t_0) = y_A(t_0) \neq 0, \qquad (22a)$$

$$\dot{y}_C(t_0) = \dot{y}_A(t_0), \qquad (22b)$$

$$(p_2 + p_0 - 1)^2 = p_1^2(1 - A_0), \qquad (22c)$$

$$\dot{y}_C(t_0) = -a_2^{-0.5}(t_0) \left[\frac{p_2 + p_0 - 1}{p_1} + f_A(t_0)\right] y_C(t_0), \qquad (22d)$$

where $k_i$'s in Eq. (3a) for system $B$ are replaced by $p_i$'s in Eq. (20) for system $C$.

Since, $(A, B)$ and $(B, C)$ are commutative under non-zero initial conditions by hypothesis, Eqs. (4a, b) and (17a, b) are satisfied; so it follows that Eqs. (22a) and (22b) are valid. Since, $B$ and $C$ are commutative, in the commutativity conditions (4c) $y'_B(t_0)$ and $y_B(t_0)$ can be replaced by $y'_C(t_0)$ and $y_C(t_0)$ due to Eqs. (15a, b); the result is

$$\dot{y}_C(t_0) = -a_2^{-0.5}(t_0)\left[\frac{k_2+k_0-1}{k_1} + f_A(t_0)\right]y_C(t_0). \tag{23}$$

On the other hand, $\dot{y}_C(t_0)$ and $y_C(t_0)$ are related by Eq. (15d). Comparing it with Eq. (23), we write

$$-b_2^{-0.5}(t_0)\left[\frac{m_2+m_0-1}{m_1} + f_B(t_0)\right]y_C(t_0) = -a_2^{-0.5}(t_0)\left[\frac{k_2+k_0-1}{k_1} + f_A(t_0)\right]y_C(t_0). \tag{24}$$

Since, $(A,B)$ is a commutative pair, substituting the values of $a_2$ from Eq. (5a) and $f_A$ from Eq. (8a) into Eq. (24), we obtain

$$-b_2^{-0.5}(t_0)\left[\frac{m_2+m_0-1}{m_1} + f_B(t_0)\right]y_C(t_0)$$

$$= -\left(\frac{b_2}{k_2}\right)^{-0.5}(t_0)\left[\frac{k_2+k_0-1}{k_1} + k_2^{-0.5}f_B(t_0) - \frac{k_1}{2k_2}\right]y_C(t_0). \tag{25}$$

Since, $b_2(t) \neq 0, y_C(t_0) \neq 0$, we can write the above equality as

$$\frac{m_2+m_0-1}{m_1} + f_B(t_0) = k_2^{0.5}\left[\frac{k_2+k_0-1}{k_1} - \frac{k_1}{2k_2}\right] + f_B(t_0). \tag{26}$$

Finally, cancelling $f_B(t_0)$, we result with

$$\frac{m_2+m_0-1}{m_1} = k_2^{0.5}\left[\frac{k_2+k_0-1}{k_1} - \frac{k_1}{2k_2}\right], \tag{27}$$

which is due to the commutativities of $(A,B)$ and $(B,C)$ under non-zero initial conditions.

Now, to prove Eq. (22c), we proceed as follows: Using Eq. (21b), we compute

$$\frac{p_2+p_0-1}{p_1} = \frac{m_2k_2+m_2k_0+\frac{m_1k_1}{2k_2^{0.5}}+m_0-1}{m_2k_1+m_1k_2^{0.5}}. \tag{28a}$$

Solving Eq. (27) for $m_1$, we have

$$m_1 = \frac{m_2+m_0-1}{k_2^{0.5}\left(\frac{k_2+k_0-1}{k_1}-\frac{k_1}{2k_2}\right)}. \tag{28b}$$

Substituting Eq. (28b) in (28a), we proceed as

$$\frac{p_2 + p_0 - 1}{p_1} = \frac{m_2(k_2+k_0)+m_0-1+\frac{k_1}{2k_2^{0,5}}\left[\frac{m_2+m_0-1}{k_2^{0,5}\left(\frac{k_2+k_0-1}{k_1}-\frac{k_1}{2k_2}\right)}\right]}{m_2 k_1 + k_2^{0,5}\left[\frac{m_2+m_0-1}{h_2^{0,5}\left(\frac{k_2+k_0-1}{k_1}-\frac{k_1}{2k_2}\right)}\right]}$$

$$= \frac{[m_2(k_2+k_0)+m_0-1]k_2^{0,5}\left(\frac{k_2+k_0-1}{k_1}-\frac{k_1}{2k_2}\right)+\frac{k_1(m_2+m_0-1)}{2k_2^{0,5}}}{m_2 k_1 k_2^{0,5}\left(\frac{k_2+k_0-1}{k_1}-\frac{k_1}{2k_2}\right)+k_2^{0,5}(m_2+m_0-1)}$$

$$= \frac{[m_2(k_2+k_0)+m_0-1]\left(\frac{k_2+k_0-1}{k_1}-\frac{k_1}{2k_2}\right)+\frac{k_1(m_2+m_0-1)}{2k_2}}{m_2 k_1\left(\frac{k_2+k_0-1}{k_1}-\frac{k_1}{2k_2}\right)+(m_2+m_0-1)}$$

$$= \frac{\frac{k_1}{2k_2}[m_2+m_0-1-m_2(k_2+k_0)-m_0+1]+\frac{k_2+k_0-1}{h_1}[m_2(k_2+k_0)+m_0-1]}{m_2(k_2+k_0-1)+m_2+m_0-1-m_2\frac{k_1^2}{2k_2}}$$

$$= \frac{k_2+k_0-1}{k_1}\frac{m_2(k_2+k_0)+m_0-1-m_2\frac{k_1^2}{2k_2}}{m_2(k_2+k_0)+m_0-1-m_2\frac{k_1^2}{2k_2^2}} = \frac{k_2+k_0-1}{k_1}. \quad (28c)$$

Using the equality (28c) in Eq. (4c) directly yields Eq. (22c). On the other hand, when Eq. (28c) is used in Eq. (23), this equation results with the proof of Eq. (22d), so does with the completion of the proof of Theorem 2.

We now introduce an example to illustrate the results obtained in the paper and to validate the transitivity by computer simulation.

**V. Example**

To illustrate the validity of the results obtained in the previous section, consider the system $A$ defined by

$$A: y''_A + (3 + \sin t)y'_A + (3.25 + 0.25\sin^2 t + 1.5 \sin t + 0.5 \cos t)y_A = x_A, \quad (29a)$$

for which Eq. (3b) yields

$$f_A(t) = \frac{a_2^{-0.5}[2a_1 - a_2]}{4} = \frac{1[2(3 + \sin t) - 0]}{4} = \frac{6 + 2\sin t}{4}$$

$$= 1.5 + 0.5 \sin t, \tag{29b}$$

$$f'_A(t) = 0.5 \cos t. \tag{29c}$$

To check Eq. (3c), we proceed

$$A_0 = a_0 - f_A^2 - a_2^{0.5} f'_A$$

$$= 3.25 + 0.25 \sin^2 t + 1.5 \sin t + 0.5 \cos t - (1.5 + 0.5 \sin t)^2 - 0.5 \cos t$$

$$= 3.25 + 0.25 \sin^2 t + 1.5 \sin t - 2.25 - 1.5 \sin t - 0.25 \sin^2 t$$

$$= 1. \tag{29d}$$

Hence, this expression is constant, that is $A_0 = 1$. Chosing $k_2 = 1, k_1 = -2, k_0 = 0$ in Eq. (3a),

$$\begin{bmatrix} b_2 \\ b_1 \\ b_0 \end{bmatrix} = \begin{bmatrix} a_2 & 0 & 0 \\ a_1 & a_2^{0.5} & 1 \\ a_0 & f_A & 0 \end{bmatrix} \begin{bmatrix} 1 \\ -2 \\ 0 \end{bmatrix} = \begin{bmatrix} a_2 \\ a_1 - 2a_2^{0.5} \\ a_0 - 2f_A \end{bmatrix}$$

$$= \begin{bmatrix} 1 \\ 3 + \sin t - 2 \\ 3.25 + 0.25\sin^2 t + 0.5 \sin t + 0.5 \cos t \end{bmatrix}$$

$$= \begin{bmatrix} 1 \\ 1 + \sin t - 2 \\ 0.25 + 0.25 \sin^2 + 0.5 \sin t + 0.5 \cos t \end{bmatrix}. \tag{30a}$$

So, $A$ and $B$ are commutative under zero initial conditions. From Eq. (30a), we compute $f_B$ and $B_0$ by using Eqs. (5b) and (5c)

$$f_B = \frac{b_2^{-0.5}(2b_1 - b_2)}{4} = \frac{(2 + 2 \sin t)}{4} = 0.5 + 0.5 \sin t, \tag{30b}$$

$$f'_B = 0.5 \cos t, \tag{30c}$$

$$B_0 = b_0 - f_B^2 - b_2^{0.5} f'_B$$

$$= 0.25 + 0.25 \sin^2 t + 0.5 \sin t + 0.5 \cos t - (0.5 + 0.5 \sin t)^2 - 0.5 \cos t$$

$$= 0.25 + 0.25 \sin^2 t + 0.5 \sin t - 0.25 - 0.5 \sin t - 0.25 \sin^2 t$$

$$= 0. \tag{30d}$$

We check the validity of Eq. (12a) by using Eqs. (30d) and (29d):

$$B_0 = k_2 A_0 + k_0 - \frac{k_1^2}{4k_2} = 1(1) + 0 - \frac{(-2)^2}{4(1)} = 0. \tag{30e}$$

It can be checked easily by using Eqs. (29b) and (30b) that Eqs. (8a,b) are also correct.

Considering the requirements for the non-zero initial conditions at $t_0 = 0$, Eq. (4) yields

$$y_B(0) = -(1)^2 \left[ \frac{1+0-1}{-2} + 1.5 + 0.5 \sin 0 \right] y_A(0) = -1.5 y_B(0). \tag{31a}$$

Hence, for the commutativity of $A$ and $B$ under non-zero initial conditions as well, due to Eqs. (6a, b) and (31a),

$$y'_A(0) = y'_B(0) = -1.5 y_B(0) = -1.5 y_A(0). \tag{31b}$$

We now consider a third system $C$ which is commutative with $B$. Therefore, using Eq. (14c) with $m_2 = 1, m_1 = 3, m_0 = 3$, we have

$$\begin{bmatrix} c_2 \\ c_1 \\ c_0 \end{bmatrix} = \begin{bmatrix} b_2 & 0 & 0 \\ b_1 & b_2^{0.5} & 0 \\ b_0 & f_B & 1 \end{bmatrix} \begin{bmatrix} 1 \\ 3 \\ 3 \end{bmatrix}.$$

Inserting values of $b_i$'s from Eq. (30a) and value of $f_B$ from Eq. (30b) in, we have

$$\begin{bmatrix} c_2 \\ c_1 \\ c_0 \end{bmatrix} = \begin{bmatrix} 1 & 0 & 0 \\ 1 + \sin t & 1 & 0 \\ 0.25 + 0.25 \sin^2 t + 0.5 \sin t + 0.5 \cos t & 0.5 + 0.5 \sin t & 1 \end{bmatrix} \begin{bmatrix} 1 \\ 3 \\ 3 \end{bmatrix}$$

$$= \begin{bmatrix} 1 \\ 4 + \sin t \\ 4.75 + 0.25 \sin^2 t + 2 \sin t + 0.5 \cos t \end{bmatrix}. \tag{32a}$$

Eqs. (16b) and (16c) yield that

$$f_C = \frac{8 + 2 \sin t}{4} = 2 + 0.5 \sin t, \tag{32b}$$

$$f'_C = 0.5 \cos t \tag{32c}$$

$$C_0 = c_0 - f_C^2 - c_2^{0.5} f'_C$$

$$= 4.75 + 0.25 \sin^2 t + 2 \sin t + 0.5 \cos t - (2 + 0.5 \sin t)^2 - 0.5 \cos t$$

$$= 4.75 + 1.25 \sin^2 t + 2 \sin t - 4 - 2 \sin t - 0.25 \sin^2 t$$

$$= 0.75. \tag{32d}$$

One can easily check that $C_0$ and $B_0$ in Eqs. (32d) and (30d) satisfy relations in Eqs. (20a,b).

For the commutativity of $B$ and $C$ under non-zero initial conditions at time $t_0 = 0$ as well; Eq. (15) together with Eqs. (30a), (30b) and chosen values of $m_i$'s yield

$$\dot{y}_C(0) = \dot{y}_B(0) = -(1)^2 \left[\frac{m_2 + m_0 - 1}{m_1} + 0{,}5 + 0{,}5 \sin 0\right] y_C(t_0)$$

$$= -\left[\frac{1+3-1}{3} + 0{,}5\right] y_C(t_0) = -1{,}5 y_C(t_0) = -1{,}5 y_B(t_0). \tag{33}$$

Considering the transitivity property under non-zero initial conditions, the conditions of Theorem 2 are satisfied. Namely, using the chosen values of $m_i$'s and $k_i$'s, from Eq. (21b), we have $p_2 = 1, p_1 = 1, p_0 = 0$. And with $A_0 = 1$ as computed in Eq. (29d), Eq. (22c) is satisfied;

$$(1 + 0 - 1)^2 = (1)^2(1-1).$$

So does Eq. (22d)

$$\dot{y}_C(0) = -(1)^{-0.5}\left(\frac{1-0-1}{1} + 1.5 + 0.5 \sin 0\right) y_C(0) = -1.5 y_C(0).$$

The answer is obviously yes as already shown in Eq. (33).

The simulations are done for the inter connection of the above mentioned systems $A, B, C$. The initial conditions are taken as

$$y_A(0) = y_B(0) = y_C(0) = 1, \tag{34a}$$

$$\dot{y}_A(0) = \dot{y}_B(0) = \dot{y}_C(0) = -1.5 y_A(0) = -1.5 y_B(0) = -1{,}5 y_C(0) = -1.5. \tag{34b}$$

And input is assumed $40\sin(10\pi t)$. It is observed that $AB, BA$ yield the same response; $BC, CB$ yield the same response; so $CA, AC$ yield to same response. These responses are shown in Fig. 2 by $AB = BA, BC = CB, CA = AC$, respectively. Hence, transitivity property shows up as if $(A, B)$ and $(B, C)$ are commutative pairs so is $(A, C)$.

These simulations and all the subsequent ones are done by MATLAB2010 Simulink Toolbox with fixed time step of 0.02 using ode3b (Bogachi-Shampine) program; the final time is $t = 10$.

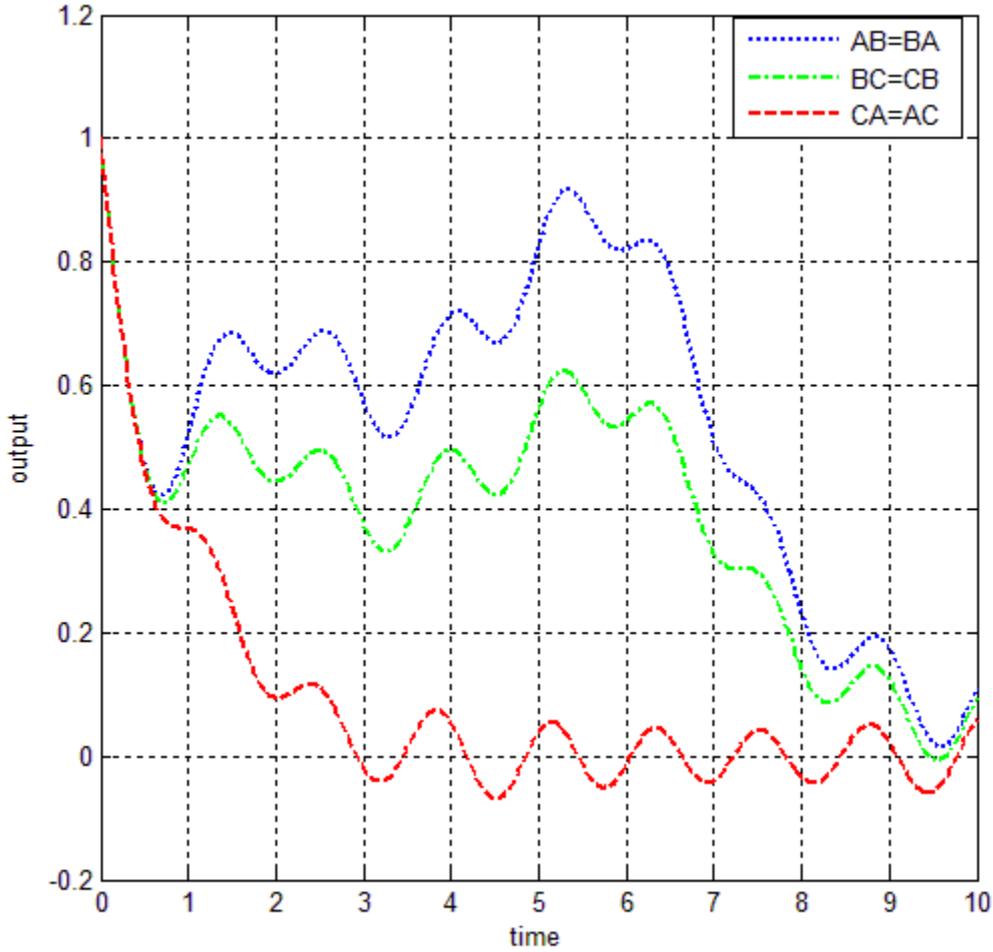

**Fig. 2:** Outputs of commutative cascade connections $AB, BA, CB, BC, CA, AC$ with nonzero initial conditions.

The second set of simulations are obtained by zero initial conditions, the conditions of Theorem 1 are satisfied by choosing $m_2 = 1, m_1 = -1, m_0 = 3$, so that $C$ is obtained from $B$ through Eq. (14) as

$$\begin{bmatrix} c_2 \\ c_1 \\ c_0 \end{bmatrix} = \begin{bmatrix} 1 & 0 & 0 \\ 1 + \sin t & 1 & 0 \\ 0.25 + 0.25 sin^2 t + 0.5 \sin t + 0{,}5 \cos t & 0.5 + 0.5 \sin t & 1 \end{bmatrix} \begin{bmatrix} 1 \\ -1 \\ 3 \end{bmatrix}$$

$$= \begin{bmatrix} 1 \\ 2 - \sin t \\ 2.75 + 0.25 sin^2 t + 0.5 \cos t \end{bmatrix}.$$

Hence, $C$ is commutative with $B$, and together with $B$ being commutative with $A$, the conditions of Theorem 1 are satisfied so that $C$ is commutative with $A$. This is observed in Fig. 3. In this figure, the responses indicated by $AB = BA, BC = CB, CA = AC$ which are all obtained by zero initial conditions validates the transitivity property of commutativity, that is Theorem 1 is valid.

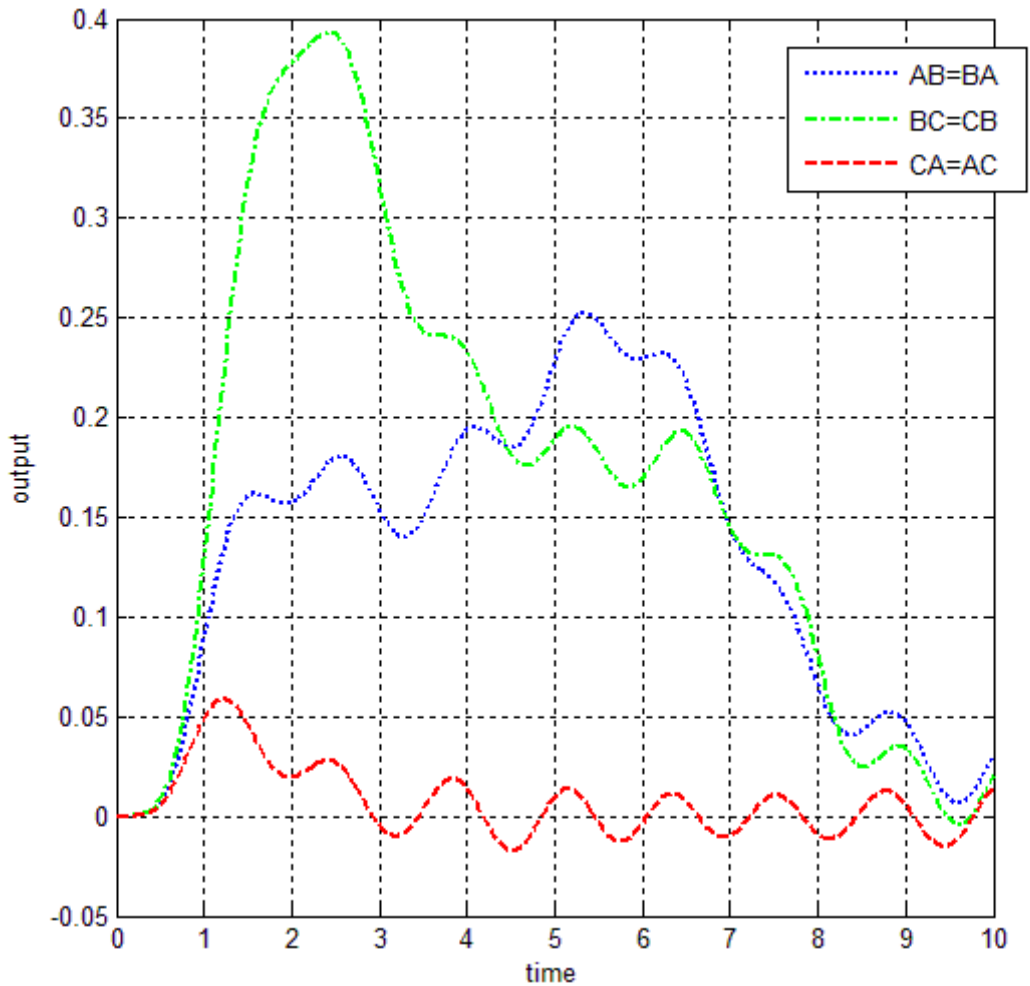

**Fig. 3:** The simulations obtained with zero initial conditions

Finally, the simulations are performed for arbitrary initial conditions $y_A(0) = 0.4$, $\dot{y}_A(0) = -0.3$, $y_B(0) = 0.2$, $\dot{y}_B(0) = -0.4$, $y_C(0) = -0.5$, $\dot{y}_C(0) = 0.5$. It is observed that $(A, B), (B, C), (C, A)$ are not commutative pairs at all, the plots AB, BA; BC, CA; CA, AC are shown in Fig. 4, respectively. However, since all systems (individually and in pairs as cascade conceded) are asymptotically stable and the effects of non-zero initial conditions die away as time proceeds, and $A, B, C$ are pairwise commutative with zero initial conditions, the responses of $AB$ and $BA$, $BC$ and $CB$, $CA$ and $AC$ approach each other with increasing time. That is commutativity property and its transitivity gets valid in the steady-state case.

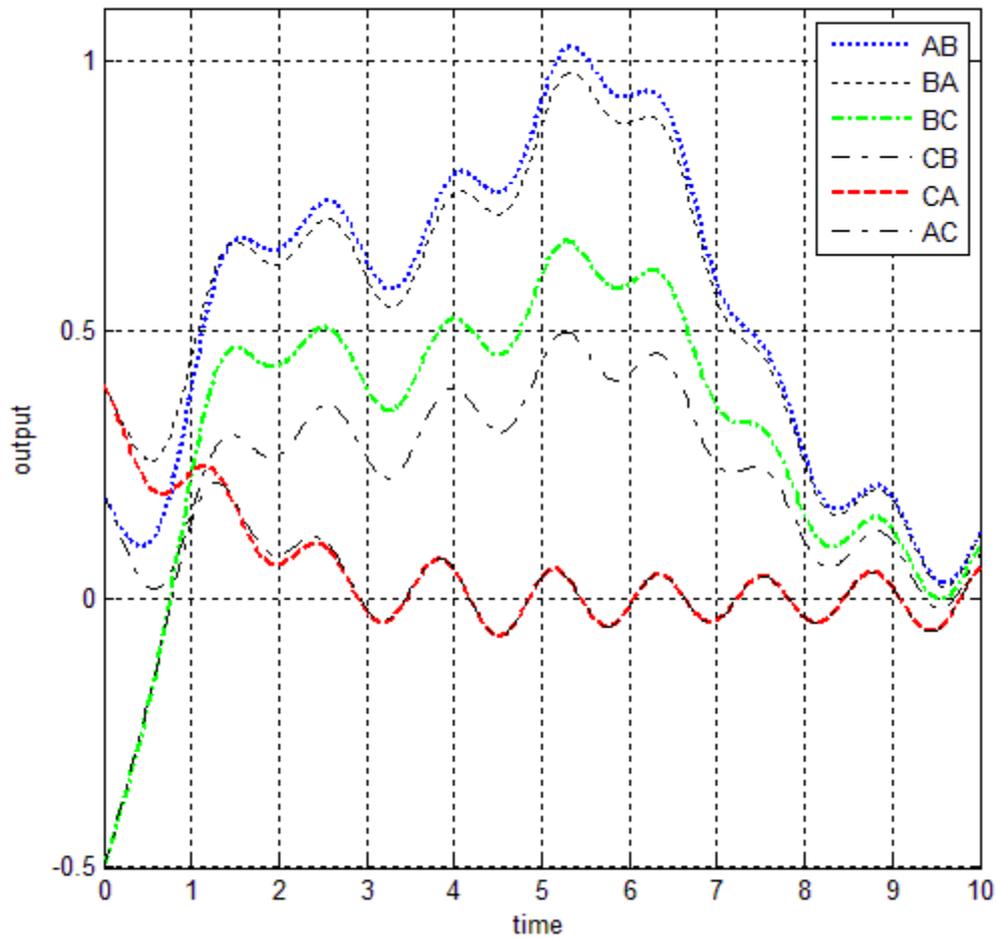

**Fig. 4:** Responses of cascade connection of Systems $A, B, C$ (which are commutative with zero initial conditions) with arbitrary initial conditions not satisfying commutativity conditions.

## VI. Conclusions

On the base of the commutativity conditions for second order linear time-varying analog systems with nonzero initial conditions, the inverse commutativity conditions are reformulated completely in the form of Lemma 1 by considering the case of non-zero initial conditions. With the obtained results, the transitivity property of commutativity is stated both for relaxed and unrelaxed cases by Theorems 1 and 2, respectively. Througout the study the subsystems considered are assumed not obtainable from each other by any feed-forword and feed-back structure, which is a case that needs special treatement due to special commutativity requirements in case of nonzero initial conditions [21].

All the results derived in the paper are well verified by simulations done by MATLAB2010 Simulink Toolbox using ode3b (Bogachi-Shampine) program.

**Acknowledgments:** This study is supported by the Scientific and Technological Research Council of Turkey (TUBITAK) under the project no. 115E952.